\documentclass[prb,twocolumn,showpacs,groupedaddress,preprintnumbers]{revtex4}%
\usepackage{amssymb}
\usepackage{amsfonts}
\usepackage{graphicx}
\usepackage{amsmath}
\usepackage{dcolumn}
\usepackage{amsmath}
\providecommand{\U}[1]{\protect\rule{.1in}{.1in}}
\begin{document}
\preprint{to appear in Physical Review B}
\title[Short Title]{Photoinduced Fano-resonance of coherent phonons in zinc}
\author{Muneaki Hase$^{1,2,3}$}
\email{mhase@bk.tsukuba.ac.jp}
\author{Jure Demsar$^{4}$}
\author{Masahiro Kitajima$^{1,2}$}
\affiliation{$^{1}$Institute of Applied Physics, University of Tsukuba, Tennodai, Tsukuba 
305-8573, Japan}
\affiliation{$^{2}$Advanced Nano-Characterization Center, National Institute for Materials
Science, Sengen, Tsukuba 305-0047, Japan}
\affiliation{$^{3}$PRESTO, Japan Science and Technology Agency, 4-1-8 Honcho,
Kawaguchi, Saitama 332-0012, Japan}
\affiliation{$^{4}$Department for Complex Matter, Jozef Stefan Institute, Jamova 39, Ljubljana, 
SI-1000, Slovenia }

\date{\today }

\begin{abstract}
Utilizing femtosecond optical pump-probe technique, we have studied transient
Fano-resonance in zinc. At high excitation levels the Fourier spectrum of the
coherent E$_{2g}$ phonon exhibits strongly asymmetric line shape, which is
well modeled by the Fano function. The Fano parameter (1/Q) was found to be
strongly excitation fluence dependent while depending weakly on the initial
lattice temperature. We attribute the origin of the Fano-resonance to the
coupling of coherent phonon to the electronic continuum, with their transition
probabilities strongly renormalized in the vicinity of the photoinduced
structural transition.

\end{abstract}
\maketitle

Fano-resonance in the linear optical spectra occurs when optical transition
paths from a given ground state to a discrete state and an overlapping
continuum of states interfere with each other\cite{Fano61}. It has been
studied in a variety of phenomena including photo-absorption in
atoms\cite{Fano61}, phonon spectra in solids\cite{Cerdeira73}, and
photo-absorption in quantum well structures\cite{Faist97}, just to mention a few.

The mechanism responsible for the Fano-resonance in phonon spectra has been
debated for decades. It has been concluded that both interband or intraband
electronic transitions can give rise to the continuum overlapping the optical
phonon energy in semiconductors, such as
Si\cite{Cerdeira73,Chandrasekhar78,Chandrasekhar80} and GaAs\cite{Nunes93}, as
well as in semimetals, like Sb\cite{Bansal86}. On the other hand, for complex
oxides, like ferroelectrics and cuprate superconductors, the continuum
originates from broad multi phonon processes\cite{Rousseau68,Mihailovic93}. In
both cases, the phonon spectra can be fitted to a Fano lineshape,
characteristic of a continuous-discrete interference:\cite{Fano61,Cerdeira73}
\begin{equation}
I(\varepsilon,Q)=\frac{(Q+\varepsilon)^{2}}{1+\varepsilon^{2}}.
\end{equation}
Here $\varepsilon=(\omega-\omega_{0}-\Delta\omega_{p})/\Gamma_{p}$, where
$\omega_{0}$ is the unperturbed frequency, $\Delta\omega_{p}$ the frequency
shift, $\Gamma_{p}$ the broadening parameter, and $Q$ is the dimensionless
asymmetry parameter. Parameters $\Gamma_{p}$ and $Q$ are defined
as\cite{Cerdeira73}
\begin{equation}
\hbar\Gamma_{p}=\pi\left\vert V_{p}\right\vert ^{2}\text{ \ \ \ \ };\text{
\ \ \ \ }\pi\Gamma_{p}Q^{2}=T_{p}^{2}/T_{e}^{2}.
\end{equation}
Here $T_{p}$ and $T_{e}$ are the probabilities for one-phonon and electronic
(or multiple-phonon) Raman scattering, respectively, and $V_{p}$ is the matrix
element of the interaction between the discrete level and continuum.

In metals there have been theoretical discussions on the Fano-interference
effects, in which both interband- and intraband-electronic Raman scattering
can contribute to the electronic continuum\cite{Klein82}. However,
experimental investigation of the Fano-interference in metals has been
inaccessible by conventional time-integrated methods like Raman
scattering.\cite{Grant73,Bolotin01} This can be attributed to either the very
fast dephasing of the electronic continuum or to screening of electron-phonon coupling.

Femtosecond (fs) real-time probes present an experimental alternative to the
conventional time-integrated methods. These techniques have an additional
advantage to conventional spectroscopy, since they utilize photoexcitation to
study both buildup and the decay of the Fano-resonance in time. Utilizing
femtosecond four-wave mixing technique Siegner \textit{et al.} studied the
dynamics of Fano-resonance in GaAs\cite{Siegner95}. Recently, transient
Fano-type antiresonance between non-equilibrium carriers and coherent LO
phonon in Si was also reported\cite{Hase03}, with the lifetime estimated to
about 100 fs. Moreover, a possibility of observing a build-up of the
Fano-resonance on a sub-fs time scale in an atom was also recently
addressed\cite{Wickenhauser05}. Thus, in the photoexcited state the coherence
between the discrete state and the continuum is expected to be short-lived,
with the relaxation dynamics governed by dephasing\cite{Siegner95}.

While photoexcited carrier dynamics in metals have been extensively studied by
fs pump-probe techniques\cite{Fann92}, only recently observation of coherent
optical phonons in metals has been reported. Using high sensitivity pump-probe
techniques, such as second harmonic generation, and transient reflectivity,
coherent optical phonons have been studied in
Gd\cite{Melnikov03,Bovensiepen04}, Cd and Zn\cite{Hase05}. However,
Fano-resonance between discrete coherent phonons and electron or
multiple-phonon continua have not yet been explored. In this paper, we present
a time-domain observation of transient Fano-resonance in metals under the
high-density optical excitation. At high perturbation densities the Fourier
transformed (FT) spectra of the coherent optical phonon exhibit strongly
asymmetric Fano-type line shapes. The line-shape depends strongly on the pump
excitation level but only weakly on the lattice temperature ($T_{l}$).

The sample studied was a single crystal of Zn with cut and polished (0001)
surface. The fs pump-probe measurements were carried out in a temperature
range from $T_{l}$ = 7 to 300 K using a cryostat. Femtosecond
pulses from a Ti:sapphire oscillator were amplified to a pulse energy of
5$\mu$J (center wavelength 800 nm and pulsewidth of 130-fs) at the repetition
rate of 100 kHz. The pump beam was focused on the sample to a spot of $\sim$
70 $\mu$m in diameter, while that of the probe beam was $\sim$ 40
$\mu$m to assure homogeneous photoexcitation. The pump power density ($F_{p}$)
was kept below 5 mJ/cm$^{2}$ to prevent sample damage \cite{damage}, while the
probe fluence was fixed at 0.1 mJ/cm$^{2}$. To record photoinduced
changes in reflectivity ($\Delta R/R$) as a function of time delay we used lock-in 
detection where the pump-beam was chopped at 2 kHz.

Figure 1(a) presents the time-resolved $\Delta R/R$ signal
obtained in Zn at $T_{l}$ = 7 K with $F_{p}$ varied between 0.68 and 4.10
mJ/cm$^{2}$. The photoinduced reflectivity transient consist of two
components\cite{Hase05}. The first is the initial transient (non-oscillatory)
response due to excitation and relaxation of nonequilibrium carriers, which
decays in a few hundred fs. The second component is the oscillatory signal due
to the generation of the coherent lattice vibration. The time period of the
observed coherent oscillation corresponds to the bulk $E_{2g}$ optical
phonon\cite{Hase05}.
\begin{figure}[ptb]
\includegraphics[width=7.0cm]{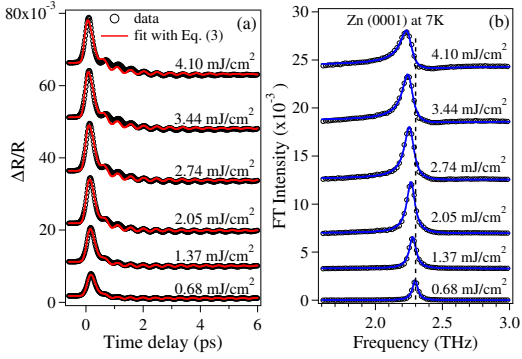}
\caption{(color online) (a) Transient
reflectivity change obtained in Zn at $T_{l}$ = 7 K at various pump fluences
and (b) their FT spectra. In both panels, open circles are the experimental
data, while solid lines present fits with Eq. (3) and Eq. (1), respectively.}
\label{Fig1}
\end{figure}

To fit the time domain data and to obtain the amplitude, frequency and the
dephasing time of the coherent optical phonon, we examined a linear
combination of a damped harmonic oscillator and a stretched exponential decay
function (STE):
\begin{equation}
\frac{\Delta R(t)}{R}=H(t)[Ae^{-\Gamma t}\cos(\omega_{p}t+\phi_{p}
)+Be^{(-t/\tau_{q})^{\alpha}}]. \label{Fit}
\end{equation}
Here $H(t)$ is the Heaviside function convoluted with Gaussian to account for
the finite time-resolution, while $A$, $\Gamma$, $\omega_{p}$ and $\phi_{p}$
are the amplitude, decay rate (damping), frequency and the initial phase of
the coherent phonon response, and $B$, $\tau_{q}$, and $\alpha$ (0 $<\alpha<$
1) are the amplitude, relaxation time and stretching parameter of the fast
electronic transient. The choice of a STE function to fit the decay of the
incoherent electronic response is merely a matter of preference and is not
associated with some underlying relaxation mechanism. Since the decay was
found to be non-exponential, as expected in this high perturbation regime, we
have chosen a STE function since it describes the incoherent response quite
accurately at all $F_{p}$ \cite{Fit}.

As shown in Fig. 1(a), at the low pump fluences the time-domain data are very
well represented by Eq. (3). The extracted frequency of the coherent optical
phonon is found to be 2.30 THz at 0.68 mJ/cm$^{2}$, in excellent agreement
with that of the $E_{2g}$ mode observed by Raman scattering\cite{Grant73}. At
high fluences, however, the coherent response cannot be fitted by a simple
damped oscillator over the entire time range. In fact, there is a significant
phase-shift of the coherent phonon at the early time delays, which cannot be
modeled by a simple damped oscillator. This discrepancy, observed only at
early time delays and in the high excitation regime, is ascribed to a coherent
correlation between the phonon and other photogenerated excitations.

\begin{figure}[ptb]
\includegraphics[width=7.0cm]{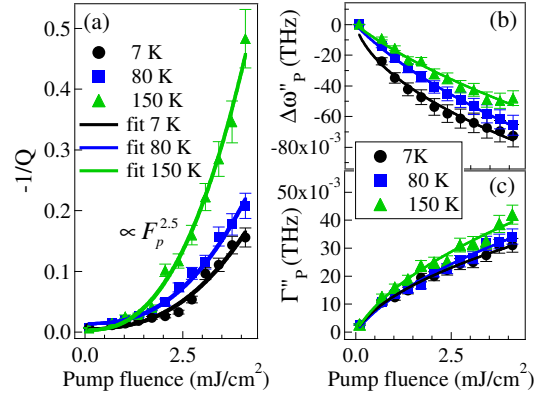}
\caption{(color online) (a) The Fano
parameter $-1/Q$, (b) the frequency shift $\Delta\omega_{p}^{"}$, and (c) the
broadening $\Gamma_{P}^{"}$ as a function of $F_{p}$ at various temperatures.
The data extracted from the low excitation intensity data (measured at $F_{p}$
= 9.2 $\mu$J/cm$^{2}$) are also plotted. The solid curves in (a) are the
fitting with $F_{p}^{2.5}$, while the solid curves in (b) and (c) show
$F_{p}^{\theta}$ behavior with $0.5<\theta<0.8$ (see text). }
\label{Fig2}
\end{figure}
The FT spectra obtained from the time domain data are shown in Fig. 1(b). They
show a red-shift of the peak frequency and broadening of the linewidth of the
$E_{2g}$ mode as $F_{p}$ increases. However, the most striking feature of the
FT spectra is an asymmetric Fano-like line shape observed at $F_{p}\gtrsim2$
mJ/cm$^{2}$. Such an asymmetric line shape has not been observed by Raman
study\cite{Bolotin01}. The spectrum can be well fitted with Fano function. The
parameters, extracted from the fit to the data with Eq. (1), are plotted in
Fig. 2 as a function of excitation level $F_{p}$ and lattice temperature. The
Fano parameter\cite{Fanoparameter} $(1/Q)$ was found to be negative, similar
to that observed for the optical phonon in n-Si using Raman
scattering\cite{Cerdeira73,Chandrasekhar78}, or the asymmetric coherent
$A_{1g}$ mode in a cuprate superconductor observed using pump-probe
technique\cite{Misochko00}. The absolute value of the Fano parameter
$\left\vert 1/Q\right\vert $ significantly increases with $F_{p}$, suggesting
that coupling between the phonon and the continuum is strongly enhanced at
high excitation densities.

As pointed out above both inter- and intra-band electronic Raman scattering
can give rise to the continuum interacting with optical phonon\cite{Klein82}.
In metals, such as Zn and Cd, interband electronic Raman scattering is
possible and would dominate total scattering efficiency\cite{Klein82}. This
process would involve interband electronic transitions along the small band
gap in the AL direction of the Brillouin zone, which is due to spin-orbit
interaction, and could be responsible for the formation of the electronic
continuum interacting with the optical phonon\cite{Bolotin01}. On the other
hand, intraband Raman scattering is also possible in metals\cite{Klein82}. Due
to the finite optical penetration depth $\delta$, the scattering wavevector
$q$ has a distribution ranging from 0 to $q_{c}\approx\delta^{-1}$. This leads
to the electronic Raman scattering with a continuum transitions over the range
of $0\leq\omega\leq q_{c}v_{F}$, where $v_{F}$ is the Fermi velocity. In Zn,
$q_{c}v_{F}\approx100$ THz, thus there is a significant overlap with the
optical phonon at $\approx$ 2.3 THz.

The phonon continuum due to two-phonon resonances (or Fermi-resonance) also
needs to be considered, since the anharmonic phonon-phonon coupling can also
be the origin of a Fano interference\cite{Mihailovic93,Kanellis86}. In Zn the
energy of the $E_{2g}$ (TO) mode is close to the difference between LA
($\approx$ 6 THz at $M$) and TA ($\approx$ 4 THz at $M$) modes (2 -2.5 THz).
Therefore, the coupling of the $E_{2g}$ mode with the two-phonon continuum is
also possible. However, the anharmonic phonon-phonon coupling depends on the
lattice temperature\cite{Kanellis86}, rather than the pump fluence. We have
analyzed the data taken at low pump fluence (9.2 $\mu$J/cm$^{2}$) from
Ref.\cite{Hase05}. The time-domain data are well fitted by Eq. (3) over the
entire temperature range. As the temperature increases from 7 K to room
temperature, the peak frequency of the FT spectra shifts to lower frequencies,
which is accompanied by broadening of the line shape. Both observations are
consistent with the anharmonic lattice effects\cite{Hase05}. However, no
asymmetry of the line is observed as in the high excitation regime. Moreover,
as seen in Fig. 2 (a), the Fano parameter strongly depends on the excitation
level and therefore on the photoexcited carrier density, but only weakly on
the initial lattice temperature.\cite{ElectronTemperature} At the low pump
fluence, $F_{p}$ = 9.2 $\mu$J/cm$^{2}$, the $-1/Q$ values over the entire
$T_{l}$-range are nearly zero, implying that the line shape is
Lorentzian\cite{Fano61}. These facts indicate that the contribution from the
two-phonon continuum is negligibly small.

\begin{figure}[ptb]
\includegraphics[width=6.0cm]{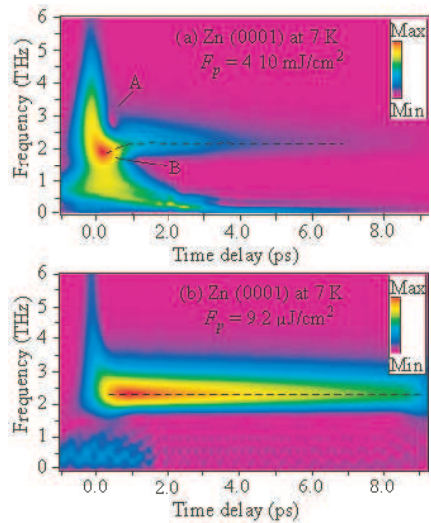}
\caption{(color online) CWT
chronograms in Zn at 7 K for two different pump fluences, (a) 4.10 mJ/cm$^{2}$
and (b) 9.2 $\mu$J/cm$^{2}$. The broad band response extending below 1 THz up
to 6 THz and decaying within a few ps in (a) represents the transient response
that is due to inter- and intra-band electronic Raman scattering, while the
dotted curves represents the peak positions of the coherent $E_{2g}$ phonon. }
\end{figure}
From the analysis presented in Figure 1, where the discrepancy between the
data and the theoretical fit is observed only at early time delays, it follows
that Fano resonance is short-lived. To further elucidate the dynamics of Fano
resonance, we utilize continuous wavelet transform (CWT) method\cite{Hase03},
to obtain the time-frequency chronograms from the time-domain data presented
in Fig. 1. Note that the incoherent part of the photoinduced reflectivity was
subtracted. As seen in Fig. 3, a signature of a Fano resonance appears at
$\approx0.5$ ps. Here both the destructive (or anti-resonance) interference at
$\approx2.7$ THz (A) and constructive interference at $\approx2.0$ THz (B) are
observed.\cite{PhononSoftening} In contrast, no anti-resonance at short time
delays is observed in the low excitation limit [Fig.3(b)] Thus, the CWT
chronogram reveals that under the high density excitation the Fano-resonance
is established on a very short time scale.

To obtain a deeper insight into the nature of the observed
excitation intensity dependence of the Fano interference we analyze the
fluence dependence of the parameters $-1/Q$, $\Delta\omega_{p}$ and
$\Gamma_{p}$ extracted from the fit of the FT data at various temperatures and
excitation fluences. To differentiate between the effects of temperature and
the excitation intensity on $\Gamma_{p}$ and $\Delta\omega_{p}$ we plot in
Figs. 2(b) and 2(c) only the intensity dependent parts $\Delta\omega_{p}^{"}$
and $\Gamma_{P}^{"}$. Here $\Delta\omega_{p}^{"}=\Delta\omega_{p}-\Delta
\omega_{p}^{0}(T)$, where $\Delta\omega_{p}^{0}$ is the value of frequency
shift in the limit of $F_{p}\rightarrow0$, extracted from the data in Ref.
\cite{Hase05}. Similarly $\Gamma_{P}^{"}=\Gamma_{p}-\Gamma_{p}^{0}(T)$, where
$\Gamma_{P}^{0}$ is the value of $\Gamma_{p}$ in the limit of $F_{p}
\rightarrow0$, and $\Gamma_{P}^{0}$ is determined from the fit to the data
obtained at $F_{p}=9.2\mu$J/cm$^{2}$.

Both $\Gamma_{P}^{"}$ and $\Delta\omega_{p}^{"}$ show a slight sublinear
dependence on $F_{p}$, $\Gamma_{P}^{"}(\Delta\omega_{p}^{"})\propto
F_{p}^{\theta}$, where $0.5<\theta<0.8$ depending slightly on temperature [see
solid lines in Fig. 2 (b) and (c)]. The Fano parameter $-1/Q$ also shows a
power law excitation intensity dependence $\left\vert 1/Q\right\vert $
$\propto$ $F_{p}^{\eta}$ and can be at all temperatures well fit by
$\eta\approx2.5$ over the wide range of excitation fluences [see Fig. 2(a)].
From the available data it is impossible to pinpoint the microscopic origin of
the photoinduced Fano interference in Zn observed in the high excitation
regime. However, some conclusions can be drawn from the observed fluence
dependence of $\left\vert 1/Q\right\vert $, $\Gamma_{P}^{"}$ and $\Delta
\omega_{p}^{"}$. Since both $\Gamma_{P}^{"}$ and $\Delta\omega_{p}^{"}$ are
roughly proportional to the square of the electron-phonon matrix element
\cite{Cerdeira73,Chandrasekhar78}, the fluence dependence of the two suggests
that photoexcitation results in an increase of the electron phonon matrix
element, where $\left\vert V_{p}\right\vert ^{2}\varpropto F_{p}^{\theta}$.
However, changes in the electron-phonon matrix element itself cannot account
for the dramatic excitation dependence of the asymmetry parameter - see Eq. 2
\cite{Note1}. In order to account for the strong excitation dependence of
$\left\vert 1/Q\right\vert $, it follows that $T_{p}^{2}/T_{e}^{2}$ should
also be strongly excitation fluence dependent. Since in the absence of changes
in the electronic and lattice structure both $T_{p}^{2}\ $and $T_{e}^{2}$
should be $F_{p}$ independent, the strong fluence dependence of $T_{p}
^{2}/T_{e}^{2}$, which follows from the strong $F_{p}$ dependence of the
asymmetry parameter, suggests a pronounced changes in the electronic structure
upon photoexcitation. Indeed, based on the published values of the absorption
coefficient, reflectivity, and plasma frequency of Zn, it follows that at
$F_{p}=5$ mJ/cm$^{2}$ photoinduced carrier density (the number density of
initially created e-h pairs) is $\approx7\%$ of the total carrier density.
This number is comparable to that of Ref. \cite{Guo00}, where photoinduced
carrier density of 15 \% resulted in band structure collapse and the
solid-to-liquid structural phase transition in photoexcited Al. We should
further note, that very similar saturation of the photoinduced reflectivity is
observed at $F_{p}\gtrsim2$mJ/cm$^{2}$ in Zn [see Fig. 1 (a)] as observed in
Al \cite{Guo00}, supporting the idea that the observed strong $F_{p}$
dependence of the Fano parameter may be due to strong renormalization of the
probabilities for one-phonon and electronic Raman scattering under high
density perturbations in the vicinity of the solid-to-liquid structural transition.

In summary, the transient photoinduced Fano-resonance between coherent phonon
and electron continuum in Zn have been observed under the high density
excitation. As the excitation fluence increases, the FT spectra exhibited
strongly asymmetric phonon lines, which were well modeled by the Fano line
shape. The broadening $\Gamma_{p}^{"}$ and the frequency shift $\Delta
\omega_{p}^{"}$ are shown to depend on the excitation level, which is
attributed to the photoinduced increase in the electron-phonon matrix element.
However, the strong fluence dependence of the Fano parameter\textbf{ $1/Q$}
implies also strong changes in the ratio of probabilities for phonon and
electronic Raman scattering. This observation is attributed to the strong
perturbation level close to the photoinduced melting.

The authors acknowledge O. V. Misochko for helpful comments. 
This work was supported in part by KAKENHI -16032218 from MEXT, Japan.


\begin{thebibliography}{99}
\bibitem {damage}At excitation levels above 5 mJ/cm$^{2}$ melting of the
sample was observed at 300 K.
\bibitem {Fit}Fano lineshape is observed no matter what kind of function is
used to subtract the incoherent response (e.g. double exponential, stretch and
so on), and the observed fluence dependence of Fano parameter is qualitatively
the same in any case.
\bibitem {Fanoparameter}It is convenient to study the evolution of the Fano
line shape in terms of the Fano parameter $(1/Q)$ where in the limit of $1/Q$
$\rightarrow$ 0, the Fano line shape becomes Lorentzian, indicating the
coupling strength is negligibly small.
\bibitem {ElectronTemperature}At the pump fluence of 4.10 mJ/cm$^{2}$, the
final electronic temperature is estimated to be $\sim$5500 K and depends
marginally on the initial temperature $T_{l}$. On the other hand, the lattice
temperature $T'_{l}$ after electron-phonon thermalization reaches 
$\sim$500 K, where the lattice heating $T'_{l}$ being dependent on $T_{l}$. 
However, since the Fano-resonance is observed within 1 ps (see
Fig. 3), contribution of the lattice heating, which arises after a few ps, to
the Fano-resonance is negligibly small.
\bibitem {PhononSoftening}The CWT spectra exhibit phonon softening at 0.5 ps,
which might be due to lattice anharmonicity and/or the plasma-induced
softening. The existence of the dip, however, can be explained neither by the
anharmonicity nor the plasma-induced softening, both of which only broaden the
phonon spectra toward low frequency side.
\end{thebibliography}
\end{document}